\documentclass[onecollarge]{svjour2}                     
\smartqed  
\usepackage{graphicx}
\usepackage{mathptmx}      
\usepackage{amsfonts}
\usepackage{amssymb}
\usepackage{amsmath}
\usepackage{bm} 
\usepackage{bbm}
\usepackage{slashed}
\usepackage{psfrag}
\journalname{Few-Body Systems}
\begin{document}

\title{Scalar-particle self-energy amplitudes and confinement  in Minkowski space\footnote{Dedicated to Professor Henryk Witala at the occasion of his 60th birthday}}


\author{Elmar P. Biernat \and Franz Gross \and Teresa Pe\~na \and Alfred Stadler}


\institute{E. P. Biernat \at CFTP (Centro de F\'isica Te\'orica de Part\'iculas), 
Instituto Superior T\'ecnico, UTL (Universidade T\'ecnica de Lisboa), 
Av. Rovisco Pais, 1049-001 Lisboa, Portugal
\\  \email{elmar.biernat@ist.utl.pt}
\and
  F. Gross \at
Thomas Jefferson National Accelerator Facility, Newport News, VA 23606, USA
 \\
and College of William and Mary, Williamsburg, VA 23187, USA
\and
T. Pe\~na \at Departamento de F\'isica and CFTP (Centro de F\'isica Te\'orica de Part\'iculas),
Instituto Superior T\'ecnico, UTL (Universidade T\'ecnica de Lisboa), 
Av. Rovisco Pais, 1049-001 Lisboa, Portugal
\and
A. Stadler\at
   Departamento de F\'isica da Universidade de \'Evora, 7000-671 \'Evora, Portugal\\
  and Centro de F\'isica Nuclear da Universidade de Lisboa, 1649-003 Lisboa, Portugal
}


\maketitle

\begin{abstract}
We analyze the analytic structure of the Covariant Spectator Theory (CST) contribution to the self-energy amplitude for a scalar particle in a $\phi^2\chi$ theory. 
To this end we derive dispersion relations in 1+1 and in 3+1 dimensional Minkowski space. The divergent loop integrals in 3+1 dimensions are regularized using dimensional regularization. We find that the CST dispersion relations exhibit, in addition to the usual right-hand branch cut, also a left-hand cut. The origin of this \lq\lq spectator'' left-hand cut can be understood in the context of scattering for a scalar $\phi^2\chi^2$-type theory. If the interaction kernel contains a linear confining component, its contribution to the self-energy vanishes exactly.


\end{abstract}

\section{Introduction}

The energy spectrum and internal structure of mesons are  influenced by both the Dirac structure of the effective confinement  interaction and the form of the low energy effective one-gluon-exchange interaction, both of great theoretical interest.   Knowledge of meson spectrum and meson wave functions allows us to unfold information about these effective interactions, and is therefore of fundamental importance to hadronic physics.  Knowing the wave functions also permits the calculation of decay rates and meson transition form factors, both of which play an important role in a large variety of reactions, including hadronic contributions to the anomalous magnetic moment of the muon \cite{Jegerlehner:2009ry} that is being investigated for possible signs of 
physics beyond the Standard Model, and in the production of lepton pairs whose precise knowledge is mandatory in the search for the quark-gluon plasma \cite{friman2011cbm}.  This strong interest has lead to the development of current and future experimental programs to measure meson spectroscopy at Jefferson Lab (CLAS12 and GLUEX programs), CERN (COMPASS), and GSI (PANDA at FAIR). Although the principal goal of these experiments is to find ``exotic'' mesons, such as glueballs and hybrids, there is still much to be learned about mesons that can be described as pure $q\bar{q}$ states. 

Our goal is to construct a unified and consistent relativistic model for mesons as  $q\bar{q}$ bound states that can be applied to both heavy and light mesons. The model should satisfy a number of important requirements: (i) the formalism used should be covariant, which is essential for the description of systems composed of light quarks; (ii) the quarks should be strictly confined; (iii) the structure of the constituent quarks themselves, i.e., their self-interaction, should be described consistently through the same confining interaction that acts between pairs of quarks; and (iv) the model should reflect the requirements of chiral symmetry (i.e. when the bare mass of a quark $q_i$ approaches zero, a massless  $q_i\bar{q}_i$  bound state must emerge).

In regard to the covariance property, an important difference between our approach and the very successful Bethe-Salpeter Dyson-Schwinger calculations \cite{Maris:2003vk} is that the latter use a Euclidean metric whereas we stay in Minkowski space. Performing a Wick rotation \cite{Wick:1954eu} of the relative-energy component to the imaginary axis, and thereby effectively turning the physical Minkowski into a Euclidean metric, has the computational advantage that the propagator singularities located on the real axis are avoided. This method thus simplifies the calculation of binding energies. However, the Bethe-Salpeter bound-state amplitudes are obtained for unphysical energy components. For the calculation of electromagnetic form factors, the Bethe-Salpeter amplitude has to be integrated over physical loop momenta, but the reverse Wick rotation crosses singularities of the integrand. In addition, for non-zero momentum transfer, the Bethe-Salpeter amplitude has to be known in a frame with non-zero total 
momentum. In principle it can be obtained through a boost from the rest-frame amplitude. However, if the rest-frame amplitude is known only for imaginary relative-energy values, the boost parameter becomes complex, and performing the boost actually requires knowledge of the amplitude in the whole complex plane. It is not clear how the required analytic continuation could be performed in a reliable manner \cite{Zuilhof:1980ae,Carbonell:2010tz,Carbonell:2008tz}. These difficulties are often ignored, but Refs.\ \cite{Carbonell:2010tz,Carbonell:2008tz} showed in simple model calculations that significant differences between form factors calculated in  Euclidean and Minkowski metric can be observed.

A different way to deal with the difficulties of the  Bethe-Salpeter equation is to simplify it without leaving Minkowski space. A number of so-called quasi-potential approaches were constructed for that purpose. They all have in common that the dimension of internal loop integrations is reduced from four to three, which makes the resulting equations much easier to manage. We use the Covariant Spectator Theory (CST) \cite{Gro69,Gro82,Gro82b}, which has been successfully applied in many different few-body systems (for a short recent review see Ref.\ \cite{Sta11}). It was also shown in model calculations with a simple interaction between two scalar particles, for which it was possible to solve the Bethe-Salpeter equation including a complete kernel of ladder and crossed ladder diagrams in a Feynman-Schwinger representation, that the binding energy obtained with the CST equation is actually closer to the exact result than the Bethe-Salpeter equation in ladder approximation (the form that is almost always used 
in applications) \cite{Nieuwenhuis:1996mc}.

In a consistent theory of mesons, the constituent quark mass should be related to the quark-antiquark interaction. In other words, the constituent quark mass should be calculated from a bare mass and the dressing generated by the same interaction. For the case of the pion, it was shown analytically that in the chiral limit of vanishing bare quark mass, the two-body CST equation yields a massless pion solution, while the quarks acquire a non-zero dressed mass through dynamical chiral symmetry breaking. However, in the numerical calculations performed so far, the constituent quark mass has been treated as a constant \cite{Gross:1991te,Gross:1991pk,Gross:1994he} or as a phenomenological function not related to the kernel \cite{Savkli:1999me}. 

From the structure of the two-body bound-state equation, which will be introduced in Sec.\ \ref{sec:model-overview}, one can see that the quark mass, which is a function of the quark four-momentum squared, appears inside a loop integral. Thus, we have to be able to calculate it  for arbitrary---including negative---values of its argument. But before we embark on this somewhat ambitious program, we will study a simpler case, namely that of a scalar particle dressed by the exchange of another scalar particle and an effective linear confining interaction. A thorough understanding of the properties of the scalar quark mass function should provide useful information also about the fermionic quark mass function. In particular, it could yield constraints on the interaction models with confinement to be used in the realistic case of quarks as fermions.

This paper is organized in the following way: after this Introduction, Sec.\ \ref{sec:model-overview} describes briefly the main features of our CST quark model. In Sec.\ \ref{sec:scalar1+1}, dispersion relations for the self-energy of a scalar particle dressed through another scalar particle of smaller mass in a $\phi^2\chi$ theory are derived in 1+1 dimensions and compared to the dispersion relations of a four-point (``bubble'') amplitude in the corresponding $\phi^2\chi^2$ theory. These calculations are extended to 3+1 dimensions in Sec.\ \ref{sec:scalar3+1}. In Sec.\ \ref{sec:scalar-conf} we consider the self-energy for linear confinement in 3+1 dimensions, and in Sec.\ \ref{sec:conclusions} we draw our conclusions.

\section{Quark-antiquark bound-state equation and quark self-energy}
\label{sec:model-overview}

The development of a covariant model of mesons as bound states of constituent quark-antiquark ($q\bar{q}$) pairs in the framework of the Covariant Spectator Theory (CST) was initiated in a series of papers \cite{Gross:1991te,Gross:1991pk,Gross:1994he,Savkli:1999me}. In this section we will briefly review the most important features of this model and discuss the aspects we plan to improve.  We begin the discussion by considering the realistic case of two Dirac particles, and later specialize to scalar particles.

\begin{figure}[tbp]
\begin{center}
\includegraphics[width=0.7\textwidth]{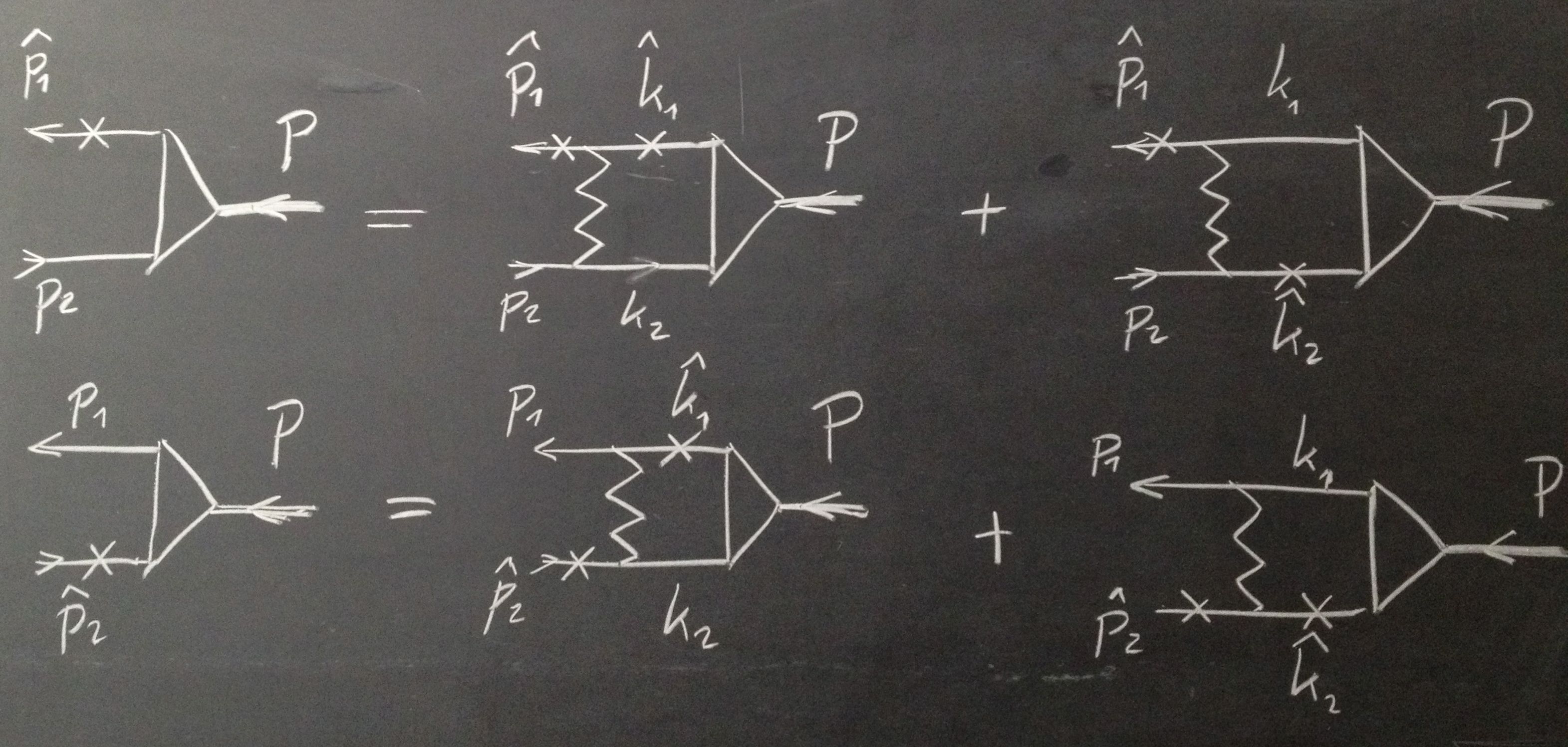}
\caption{The coupled CST equations for the $q\bar{q}$ bound-state vertex function (triangle). The zigzag line represents the kernel, which in this paper is the sum of a confining interaction and the exchange of another particle.  A cross on a quark line indicates that the particle is on mass shell.}
\label{Fig:qqbarCST}
\end{center}
\end{figure}

The general structure of the CST equations for the $q\bar{q}$ bound-state vertex function $\Gamma$ is displayed graphically in Fig.\ \ref{Fig:qqbarCST}. 
The interaction kernel $V$ is an operator in the Dirac space of the two quarks. It can be written in the general form 
\begin{equation}
V=\sum_i {\cal O}_1^{(i)} \otimes {\cal O}_2^{(i)} V^{(i)} \, ,
\label{eq:qqbar-kernel}
\end{equation}
where 
${\cal O}_j^{(i)}$ refers to a Dirac matrix of type $i$ in the space of particle $j$, and $V^{(i)}$ describes the momentum dependence of that particular type of interaction (generally confinement plus particle exchanges). 
Assuming that particle 1 is the quark with four-momentum $p_1$ and particle 2 the antiquark with four-momentum $p_2$, the algebraic form of the two-body equation is 
\begin{align}
\Gamma_1(\hat{p}_1,p_2) & = \sum_i \int \frac{\mathrm d^3 k_1}{2E_{k_1}(2\pi)^3} 
{\cal O}_1^{(i)} (m+ \hat{\slashed{k}}_1)  \Gamma_1(\hat{k}_1,k_2) S(k_2) {\cal O}_2^{(i)} V^{(i)}(\hat{p}_1,\hat{k}_1)
 \nonumber \\
 & + 
 \sum_i \int \frac{\mathrm d^3 k_2}{2E_{k_2}(2\pi)^3} 
{\cal O}_1^{(i)}  S(k_1)  \Gamma_2(k_1,\hat{k}_2)  (m+ \hat{\slashed{k}}_2) {\cal O}_2^{(i)} V^{(i)}(\hat{p}_1,k_1)
 \nonumber \\
\Gamma_2(p_1,\hat{p}_2)  &= \sum_i \int \frac{\mathrm d^3 k_1}{2E_{k_1}(2\pi)^3}
{\cal O}_1^{(i)} (m+ \hat{\slashed{k}}_1)  \Gamma_1(\hat{k}_1,k_2) S(k_2) {\cal O}_2^{(i)} V^{(i)}(p_1,\hat{k}_1)
 \nonumber \\
 & + 
 \sum_i \int \frac{\mathrm d^3 k_2}{2E_{k_2}(2\pi)^3} 
{\cal O}_1^{(i)}  S(k_1)  \Gamma_2(k_1,\hat{k}_2)  (m+ \hat{\slashed{k}}_2) {\cal O}_2^{(i)} V^{(i)}(p_1,k_1)
\, .
\label{eq:qqbarCST}
\end{align}
We use a notation in which a hat over a particle's momentum explicitly indicates that it is on mass shell, e.g., $\hat{p}_i^2=m^2$, whereas $p_i$ is in general off mass shell.
Furthermore, $P=\hat{p}_1+p_2=p_1+\hat{p}_2$ is the total momentum of the meson, $m$ the dressed quark mass, and $E_k \equiv (m^2+{\bm k}^2)^{1/2}$. The propagator for an off-shell quark with four-momentum $p$ is
\begin{equation}
S(p)=\frac{1}{m_0-\slashed{p}+\Sigma(p)-i\epsilon}=Z(p^2) \frac{M(p^2)+\slashed{p}}{M^2(p^2)-p^2-i\epsilon} \, ,
\end{equation}
where the dressed quark mass function $M(p^2)$ and the renormalization factor $Z(p^2)$ are related to the quark self-energy $\Sigma(p)$ through
\begin{align}
\Sigma(p) & = A(p^2)+\slashed{p} B(p^2) \, , \\
M(p^2) & = Z(p^2) \left[ m_0 + A(p^2) \right] \, , \\
Z(p^2) & = \left[ 1-B(p^2) \right]^{-1} \, ,
\end{align}
and $m_0$ is the bare quark mass. The dressed mass $m$ is the solution to the equation $M(m^2)-m=0$.

It is one of the main features of the CST of a two-body system that in intermediate states the heavier particle is placed on its positive-energy mass shell. Since we are particularly interested in obtaining solutions for light mesons made of up and down quarks which are treated as equal-mass particles, we have to generalize this prescription in a way that restricts either particle on mass shell with equal weight. This leads to the system of two coupled equations of Fig.\ \ref{Fig:qqbarCST} and Eq.\ (\ref{eq:qqbarCST}). The two-channel formulation is also necessary to ensure consistency in the limit when $m_0\to0$, necessary for a study of the chiral limit \cite{Gross:1991te}.
For the description of heavy-light quark systems, the system of coupled equations can be reduced to a single equation, consisting of the first equation of Fig.\ \ref{Fig:qqbarCST} with only the first term on the right-hand side. 

For a realistic model of $q\bar{q}$ bound states, the kernel must include a confining interaction. Nonrelativistic potential models and lattice QCD studies suggest that the strength of this confining interaction increases linearly with the distance between the quarks, but there is no general consensus about its Lorentz structure. 

It is not obvious how a linear potential can be used in momentum space, since its Fourier transform does not exist. Nevertheless, Ref.\ \cite{Gross:1991te} introduced a limiting procedure which allows an exact solution of bound states in a nonrelativistic linear potential in momentum space, and which can be generalized to the relativistic case, although such generalizations are clearly not unique and require additional constraints.
According to this method, the momentum-space version of the nonrelativistic linear potential ${\tilde V}(r)=\sigma r$ can be written in the form
\begin{equation}
V_\mathrm L({\bm q}) = \lim_{\epsilon \rightarrow 0} \left[ V_\mathrm{A}({\bm q}) -(2\pi)^3\delta^{(3)}({\bm q})\int \frac{\mathrm d^3 q'}{(2\pi)^3} V_\mathrm A({\bm q}') \right] \, ,
\label{eq:VL}
\end{equation}
where $\bm q$ is the momentum transfer, and
\begin{equation}
V_{\mathrm A}({\bm q}) = - \frac{8\pi\sigma}{({\bm q}^2+\epsilon^2)^2} \, .
\label{eq:VA}
\end{equation}
The subtraction term proportional to the delta function in (\ref{eq:VL}) is crucial. The Fourier transform of $V_A$ alone back into $r$-space diverges as $\epsilon \rightarrow 0$, and the subtraction term exactly cancels the singular part. A straightforward covariant generalization is the simple substitution ${\bm q}^2 \rightarrow -q^2$, which guarantees that a linear confining potential is obtained in the nonrelativistic limit. This has advantages, but also disadvantages, which is why other forms have also been explored \cite{Savkli:1999me}. Which form is the most suitable to be used in the relativistic kernel is still under investigation.

In past CST calculations, the linear potential was chosen to be purely scalar, and a constant vector interaction was added to adjust the mass scale. Additional interaction terms can be included if needed, but the exact form of the kernel will eventually be obtained from a fit to the meson spectrum. 

However, the choice of allowed interactions in the kernel is constrained. For instance, chiral symmetry requires that in the chiral limit, i.e., for vanishing bare quark mass, the pion must become massless. The dressed mass of the constituent quarks is generated through the dynamical breaking of chiral symmetry. In the CST quark model, this can be implemented  in a way that is analogous to the Nambu--Jona-Lasinio (NJL) mechanism \cite{Nambu:1961tp}, where chiral symmetry is dynamically broken through point-like fermion self-interactions. In our case, the self-interaction is of a more general kind, because it also includes confinement.

In fact,  when the quark mass function is calculated from the self-energy generated by a certain class of relativistic $q\bar{q}$ kernels, in the chiral limit the $q\bar{q}$ equations automatically have a zero-mass bound-state solution \cite{Gross:1991te}. In general, this can only occur with relativistic kernels that are invariant under chiral rotations. Remarkably, it also holds for scalar confining interactions. Although a scalar term breaks chiral symmetry, in the chiral limit it completely decouples from the equations and is therefore an allowed component of the kernel.

As already mentioned in the Introduction, we want to eliminate a lack of consistency in previous calculations of the structure of light mesons in the CST, where the dressed quark mass was treated as a constant parameter. In this work, we calculate the self-energy of scalar quarks which allows us to substitute this constant quark mass by the correct mass function. 

A convenient form to represent the mass function is in terms of a dispersion integral which reveals its analytic structure. 
Here we have to address a more fundamental question that arises in this context: what kind of dispersion relations do the self-energy amplitudes calculated in the CST framework satisfy? This is an interesting question, because by selecting certain poles in the complex relative-energy plane and excluding others, the analytic structure of the corresponding amplitudes is surely altered, and it is important to understand the consequences of these changes.  

In the following, we will begin by deriving the dispersion relations for the self-energy amplitude for the simple case of one scalar particle dressed by another scalar particle with a different mass.  Before considering the 3+1 dimensional case, it is useful to analyze the problem first in 1+1 dimensions where one can obtain analytic results for both the Feynman and CST amplitudes without any necessity to regularize divergent loop integrals.

\section{Two scalar particles in 1+1 dimensions}
\label{sec:scalar1+1}

The aim of this section is to demonstrate how the self-energy amplitude is derived within the CST framework. We can compare our results with another approximation used  by Becher and Leutwyler in Ref.~\cite{Bec99} in the context of heavy-baryon chiral perturbation theory, and we apply some of the methods to derive dispersion relations described there.

\begin{figure}\begin{center}

  \includegraphics[width=0.4\textwidth]{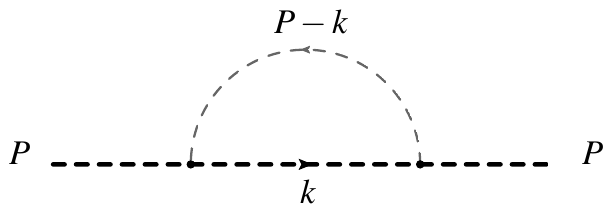}
\caption{The self-energy diagram.}
   \label{fig:selfenergy}   
   \end{center}
\end{figure}            

The self-energy corresponds to the one-loop graph shown in Fig.~\ref{fig:selfenergy}.
At first, in order to keep the calculations simple and to avoid singularities emerging from divergent integrals, we consider the scalar-loop diagram involving two scalar particles of masses $M$ and $m$ with $M>m$ in two-dimensional Minkowski spacetime.\footnote{We use the standard convention for the metric tensor $\mathrm{diag} (\mathrm g_{\mu\nu})=(1,- 1)$.} This simple example serves as a review of the basic concepts and ideas. Then, in the next section we turn to the more intricate case of four space-time dimensions where the scalar-loop integral diverges, which we regularize through dimensional regularization. 
\subsection{Dispersion relation}
\label{sec:Disprel}
This subsection reviews the dispersion relation for the self-energy diagram in two dimensions and serves as a preparation for the subsequent sections. From Fig.~\ref{fig:selfenergy} the familiar Feynman scalar-loop integral is given by 
\begin{equation}\label{eq:B(s)1}
\Sigma(s)=\mathrm i\int\frac{\mathrm d^2 k}{(2\pi)^2}\frac{1}{A_1A_2}
\end{equation} with 
\begin{eqnarray}\label{eq:A1andA2}
A_1&=&M^2-k^2-\mathrm i \epsilon_1\,,\nonumber \\
A_2&=&m^2-(P-k)^2-\mathrm i \epsilon_2
\end{eqnarray} 
where $k^\mu=(k^0,k^1)$ is the internal two-momentum and $P^\mu=(P^0,P^1)$ is the total two-momentum with Mandelstam $s=P^2$.  The values of coupling constants are irrelevant in this context and thus are set to 1 throughout this work. The \lq $\mathrm i \epsilon$' prescriptions in the denominators of the propagators impose the boundary condition that positive-energy states propagate forward and negative-energy states backward in time. Note that we have equipped the two masses with \textit{different} imaginary parts whose zero limits are taken \textit{one after the other}. Although this requirement is not needed for now it will become relevant in the next subsection.

The product of the two propagators in Eq.~(\ref{eq:B(s)1}) can be rewritten using the standard Feynman parametrization as
\begin{eqnarray}\label{eq:A1A2}
\frac{1}{A_1A_2}=\int_0^1 \frac{ \mathrm dz}{(A_2z+A_1(1-z))^2}\quad.
\end{eqnarray}
Performing the integrations over $\mathrm d^2k$ gives 
\begin{equation}\label{eq:B(s)2}
\Sigma(s)=-\frac{1}{4\pi}\int_0^1 \frac{\mathrm dz}{C(z,s)}
\end{equation}
with 
\begin{equation}\label{eq:C}
C(z,s)=z(1-z)\left[\frac{m^2}{1-z}+\frac{M^2}{z}-s-\mathrm i\left(\frac{\epsilon_2}{1-z}+\frac{\epsilon_1}{z}\right)\right]\,.
\end{equation}

\begin{figure}\begin{center}

  \includegraphics[width=0.6\textwidth]{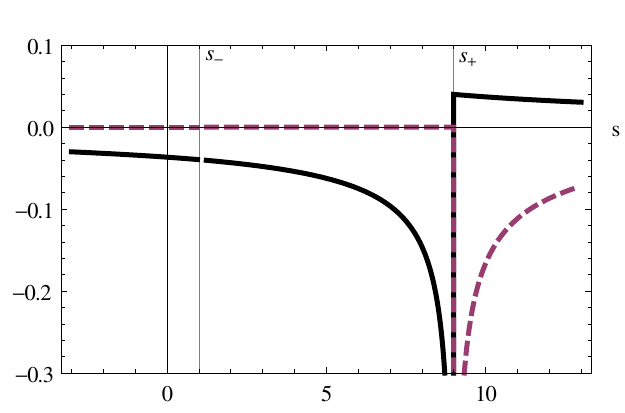}
\caption{The real (solid) and imaginary (dashed) parts of $\Sigma(s)$ for the choice $M=2$ and $m=1$.}
   \label{fig:Sigma}   
   \end{center}
\end{figure}

In the $z$-interval of integration $[0,1]$ the imaginary part in the square brackets is always negative, therefore it is safe to make the following replacement in that region without changing the \lq $\mathrm i\epsilon$' prescription:
\begin{equation}\label{eq:epsilon}
\frac{\epsilon_2}{1-z}+\frac{\epsilon_1}{z}\rightarrow\epsilon\,.
\end{equation}
The integral of Eq.~(\ref{eq:B(s)2}) is elementary and can be written as
\begin{equation}\label{eq:B(s)3}
\Sigma(s)=-\frac{\mathrm i}{2\pi\rho(s+\mathrm i \epsilon)}\left\lbrace\mathrm{arctan}\left[\eta_-(s+\mathrm i \epsilon)\right]+\mathrm{arctan}\left[\eta_+(s+\mathrm i \epsilon)\right] \right\rbrace\,
\end{equation}
where
\begin{equation}\label{eq:rho(s)2}
\rho(s)=\sqrt{(s-s_+)(s-s_-)}
\end{equation} is the familiar two-body phase space factor with
\begin{eqnarray}\label{eq:s+s-}
s_+=(M+m)^2\,, \quad s_-=(M-m)^2\, \quad
\text{
and}\quad \label{eq:etapm}
\eta_\pm(s)=-\mathrm i\,\frac{s\pm(M^2-m^2)}{\rho(s)}\,.
\end{eqnarray}
The real and imaginary parts of $\Sigma(s)$ are depicted in Fig.~\ref{fig:Sigma}. 

\begin{figure}\begin{center}

  \includegraphics[width=0.75\textwidth]{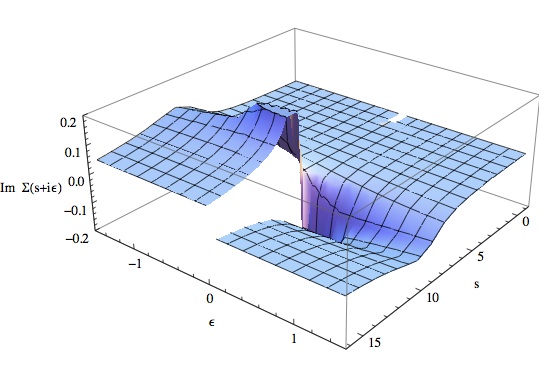}
\caption{The branch cut discontinuity of $\mathrm {Im}\Sigma(s+\mathrm i \epsilon)$ for the choice $M=2$ and $m=1$. Note that in order to display the discontinuity we have plotted $\mathrm {Im}\Sigma$ as a function in the complex $(s+\mathrm i \epsilon)$-plane (without performing the limit $\epsilon\rightarrow0$).}
   \label{fig:ImSigma3D}   
   \end{center}
\end{figure}

Note that the imaginary part of $\Sigma(s)$ is non-zero only above the threshold $s>s_+$ where it is kinematically possible that both particles are on mass shell. It has a singularity at $s=s_+$ and a discontinuity along the branch cut on the positive real axis from $s_+$ to $\infty$ as illustrated in Fig.~\ref{fig:ImSigma3D}. 
The \lq$\mathrm i \epsilon$' prescription in the arguments ensures that $\mathrm {Im}\Sigma(s)$ is always evaluated on the correct side of the cut, which is in the case at hand the upper-half complex $s$ plane.  
We proceed with our discussion by showing that the self-energy $\Sigma(s)$ satisfies a dispersion relation. In order to find the corresponding integral representation we perform a variable transformation of the $z$ integral in Eq.~(\ref{eq:B(s)2}) defined by the map
\begin{eqnarray}\label{eq:z->sprime}
s'(z)=\frac{m^2}{1-z}+\frac{M^2}{z}\,.
\end{eqnarray}
The inverse transformation involves two roots:
\begin{eqnarray}\label{eq:sprime->z}
z_\pm(s')=\frac{1}{2s'}\left[s'+M^2-m^2\pm \rho(s')\right]\,.
\end{eqnarray}
On the interval of integration $0<z<1$, the negative branch covers the interval $0<z_-<z(s_+)\equiv M/(M+m)$ and the positive branch covers the interval $z(s_+)<z_+<1$. Furthermore, each of these intervals is mapped onto the same $s'$ interval $s_+< s'<\infty$ and this interval is thus covered twice by the mapping of Eq.~(\ref{eq:z->sprime}). Written in terms of the new variable $s'$, the integral of Eq.~(\ref{eq:B(s)2}) acquires the expected dispersive form 
\begin{eqnarray}\label{eq:B(s)5}
\Sigma(s)=-\frac{1}{2\pi}\int^\infty_{s_+}\frac{\mathrm d s'}{\rho(s')}\frac{1}{s'-s-\mathrm i \epsilon }=\frac{1}{\pi} \int_{s_+}^\infty \frac{\mathrm d s'\mathrm {Im}\Sigma(s')}{s'-s-\mathrm i \epsilon } 
\end{eqnarray}
with
\begin{eqnarray}\label{eq:ImSigma}
\mathrm {Im}\Sigma(s)= -\frac{1}{2\rho(s)}\theta(s-s_+)\,.
\end{eqnarray}

We have checked numerically that the analytic expression of Eq.~(\ref{eq:B(s)3}) agrees with the dispersion integral of Eq.~(\ref{eq:B(s)5}).    

\subsection{Dispersion relation in the spectator theory}
\label{sec:speccontribution}

In this section we turn to the CST framework and derive the corresponding dispersion relation for the self-energy. Applying the CST program to the self-energy diagram means that the heavier particle of mass $M$ is placed on its positive-energy mass shell. At this point one might ask what prevents us from using the full self-energy amplitude, which was just obtained in Eqs.\ (\ref{eq:B(s)3}) and (\ref{eq:B(s)5}), to calculate the mass function for the off-shell quark. The reason is that in the chiral limit of vanishing bare quark mass the $q\bar{q}$ equation with a confining interaction kernel should have a zero-mass bound-state solution. Such a solution of Eq.\ (\ref{eq:qqbarCST}) is obtained automatically only if the quark inside the self-energy loop is on mass shell. Although this requirement is not strictly necessary for scalar particles, we nevertheless apply it here because this work serves mainly as preparation for more realistic cases, in particular of the pion where the CST constraint is 
indispensable.

An elegant procedure to work out this \lq spectator' contribution has been presented in Refs.~\cite{Gro82,Bec99,Gro04up}. The idea is to split up the product of the two propagators of Eq.~(\ref{eq:A1A2}) into two terms according to the algebraic identity
\begin{eqnarray}\label{eq:A1A2split}
\frac{1}{A_1A_2}=\frac{1}{A_1(A_2-A_1)}-\frac{1}{A_2(A_2-A_1)}\,, 
\end{eqnarray}
or, equivalently, in terms of the $z$ integration as
\begin{eqnarray}\label{eq:A1A2split2}
\int_0^1 \frac{\mathrm dz }{(A_2z+A_1(1-z))^2}=\int_0^\infty  \frac{\mathrm dz}{(A_2z+A_1(1-z))^2}-\int_1^\infty  \frac{\mathrm dz}{(A_2z+A_1(1-z))^2}\,.
\end{eqnarray}
We can design this representation in such a way that the desired spectator contribution is given entirely and only by the first term on the right-hand side of this equation by choosing an appropriate ``$\epsilon$-prescription''.
In order to see this explicitly we go to the overall rest frame, where $P=(W,0)$ with $W=\sqrt{s}$. Inserting for the propagators this contribution becomes
\begin{equation}\label{eq:Bspec1}
\Sigma_\text{CST}(s)=\mathrm i\int \frac{\mathrm d k_1\mathrm d k_0}{(2\pi)^2}\frac{1}{(E_k-k_0-\mathrm i \epsilon_1)(E_k+k_0-\mathrm i \epsilon_1)\left[m^2-M^2-s+2Wk_0-\mathrm i (\epsilon_2-\epsilon_1)\right]}\,,
\end{equation}
where $E_k=\sqrt{M^2+k_1^2}$. At this point it is obvious why we have equipped the two masses with different imaginary parts: it simply avoids the cancellation of the two imaginary parts and thus provides a well-defined \lq$\mathrm i \epsilon$' prescription of the pole in the square brackets. Otherwise, the absence of an \lq$\mathrm i \epsilon$' prescription would make the integral over $k$ ambiguous~\cite{Bec99}. In order to obtain the spectator contribution of the heavy particle on its positive-energy mass shell we assume the following \lq$\epsilon$-limiting' prescription: \textit{first} take the limit $\epsilon_1\rightarrow0$ and \textit{then} take the limit $\epsilon_2\rightarrow0$ (such that the difference $\epsilon_2-\epsilon_1$ in~(\ref{eq:Bspec1}) is always positive) and assume that $W>0$. In this case there is only one pole in the lower-half complex $k_0$ plane, the spectator pole at $k_0=E_k-\mathrm i \epsilon_1$. Now we can carry out the $k_0$ integration and we recognize the familiar CST term (
see, e.g., Ref.~\cite{Sta11}):
\begin{equation}\label{eq:Bspec2}
\Sigma_\text{CST}(s)=-\int\frac{\mathrm d k_1}{2\pi}\frac{1}{2E_k\left[e_k^2-(W-E_k)^2\right]}\,
\end{equation}
where $e_k=\sqrt{m^2+k_1^2}$. 

Note that we would have obtained the same result if we would have taken the opposite \lq$\epsilon$-limiting' prescription provided $W<0$. For the other two possible assumptions the single pole would have been in the upper half plane at $k_0=-E_k+\mathrm i \epsilon_1$, corresponding to taking the negative-energy pole of the heavy particle. Therefore, the \lq$\epsilon$-limiting' prescription together with the sign of $W$ simply fixes the position of the pole $k_0=(s+M^2-m^2)/2W$ in the complex plane. 

With the same \lq$\epsilon$-limiting' prescription as above and in an analogous manner one can show that the second term on the right hand side of Eq.~(\ref{eq:A1A2split}) corresponds to the negative-energy pole contribution of the light particle of mass $m$.

In order to find an analytic expression for $\Sigma_\text{CST}(s)$ we consider the first term on the right hand side of Eq.~(\ref{eq:A1A2split2}) and integrate over $\mathrm d^2 k$ to obtain
\begin{equation}\label{eq:Bspec3}
\Sigma_\text{CST}(s)=-\frac{1}{4\pi}\int_0^\infty  \frac{\mathrm dz}{C(z,s)}\,.
\end{equation}
This resembles the result in Eq.~(\ref{eq:B(s)2}), with the difference that the upper limit is now $\infty$ instead of 1. Before performing the $z$ integration we have to analyze the \lq$\mathrm i \epsilon$' prescriptions over the relevant $s$ regions in order to obtain the correct sign of $\mathrm {Im}\Sigma_\text{CST}(s)$. With our \lq$\epsilon$-limiting' prescription defined above the imaginary part in the square brackets of Eq.~(\ref{eq:C}) is negative for $0<z<1$ and positive for $z>1$. This, in effect, results in a sign change of the \lq$\mathrm i \epsilon$' prescription on how to treat the pole at $s'(z)=s$. Note that in the interval $0<z<1$ the function $s'(z)$ has a minimum $s_+$ at $z=M/(M+m)$ and approaches $\infty$ as $z\rightarrow 0^+$ or $z\rightarrow 1^-$. Further, in the interval $1<z<\infty$ the function $s'(z)$ has a maximum $s_-$ at $z=M/(M-m)$, approaches $-\infty$ as $z\rightarrow1^+$ and becomes $0$ as $z\rightarrow\infty$. Therefore we have a \lq$+\mathrm i \epsilon$' prescription in 
the interval $s_+<s<\infty$ and a \lq$-\mathrm i \epsilon$' prescription in the interval $-\infty<s<s_-$. We can now perform the $z$ integration in Eq.~(\ref{eq:Bspec3}) to obtain
\begin{eqnarray}\label{eq:Bspec4}
\Sigma_\text{CST}(s)=\begin{cases}-\frac{\mathrm i}{2\pi\rho(s-\mathrm i \epsilon)}\left\lbrace\frac\pi2+\mathrm{arctan}\left[\eta_+(s-\mathrm i \epsilon)\right] \right\rbrace\,,
& -\infty<s<s_-\\
-\frac{\mathrm i}{2\pi\rho(s)}\left\lbrace\frac\pi2+\mathrm{arctan}\left[\eta_+(s)\right] \right\rbrace\,,
& s_-<s<s_+\\
-\frac{\mathrm i}{2\pi\rho(s+\mathrm i \epsilon)}\left\lbrace\frac\pi2+\mathrm{arctan}\left[\eta_+(s+\mathrm i \epsilon)\right] \right\rbrace\,,
& s_+<s<\infty\quad.  
\end{cases}
\end{eqnarray}
\begin{figure}\begin{center}

  \includegraphics[width=0.75\textwidth]{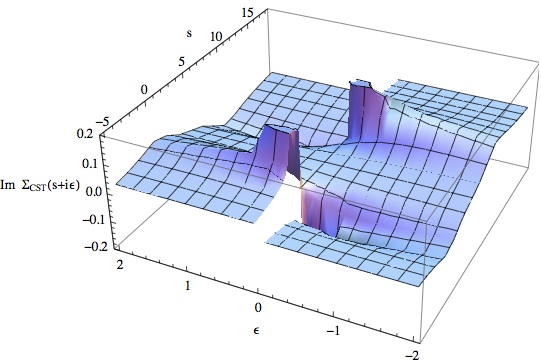}
\caption{The branch cut discontinuity of $\mathrm {Im}\Sigma_\text{CST}(s+\mathrm i \epsilon)$ for the choice $M=2$ and $m=1$.}
   \label{fig:ImSigmaCST3D}   
   \end{center}
\end{figure}            

Our next task is to write $\Sigma_\text{CST}(s)$ in a dispersive representation similar to Eq.~(\ref{eq:B(s)5}). To this end we proceed in an analogous manner as in Sec.~\ref{sec:Disprel} by writing Eq.~(\ref{eq:Bspec3}) in terms of the $s'(z)$ variable. Note that the $z$ interval $1<z<z(s_-)$ covered by the negative branch $z_-(s)$ is mapped onto the $s'$ interval $-\infty<s'<s_-$ and the interval $z(s_-)<z<\infty$ covered by the positive branch $z_+(s)$ is mapped onto $s_->s'>0$. Therefore, the interval $0<s'<s_-$ is covered twice by the map $z\rightarrow s'(z)$. As the analysis after Eq.~(\ref{eq:Bspec3}) revealed there is a \lq $-\mathrm i\epsilon$' prescription in the interval $-\infty<s<s_-$ and a \lq $+\mathrm i\epsilon$' prescription in the interval $s_+<s<\infty$. Taking this into account we obtain

\begin{eqnarray}\label{eq:Bspec5}
\Sigma_\text{CST}(s)=\frac{1}{2\pi}\left\lbrace\left[\frac12\int_{-\infty}^0+\int_{0}^{s_-}\right]\frac{\mathrm d s'}{\rho(s')}\frac{1}{s'-s+\mathrm i \epsilon }-\int^\infty_{s_+}\frac{\mathrm d s'}{\rho(s')}\frac{1}{s'-s-\mathrm i \epsilon }\right\rbrace 
\end{eqnarray}
from which we find the following expression for $\mathrm {Im} \Sigma_\text{CST}(s)$:
\begin{eqnarray}\label{eq:ImSigmaSpec}
\mathrm {Im}\Sigma_\text{CST}(s)= -\frac{1}{2\rho(s)}\left[\frac12\theta(-s)+ \theta(s)\theta(s_--s)+\theta(s-s_+)\right]\,.
\end{eqnarray}
Note that $\mathrm {Im}\Sigma_\text{CST}(s)$ vanishes in the interval $ s_-<s<s_+$, it has a discontinuity at $s=0$ and it is singular at $s=s_-$ and $s=s_+$. Further, it has two cuts along the real axis, the usual \lq right-hand cut' going from $s_+$ to $\infty$ and a \lq left-hand cut' going from $-\infty$ to $s_-$ as shown in Fig.~\ref{fig:ImSigmaCST3D}.
The numerical computation of the dispersion integral of Eq.~(\ref{eq:Bspec5}) agrees with the analytic expressions of Eq.~(\ref{eq:Bspec4}). 

A comparison of Eq.~(\ref{eq:Bspec5}) with Eqs.~(\ref{eq:A1A2split}) and~(\ref{eq:B(s)5}) reveals where the  \lq spectator left-hand cut' stems from: it corresponds to (minus) the negative-energy pole contribution of the light particle that is omitted in the spectator approximation.

\begin{figure}\begin{center}

  \includegraphics[width=0.75\textwidth]{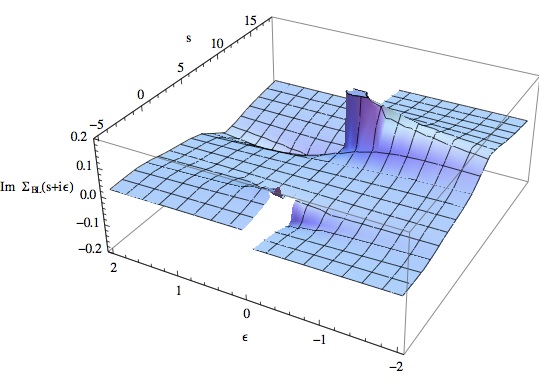}
\caption{The branch cut discontinuity of $\mathrm {Im}\Sigma_\text{BL}(s+\mathrm i \epsilon)$ for the choice $M=2$ and $m=1$.}
   \label{fig:ImSigmaBL3D}   
   \end{center}
\end{figure}

We shall now discuss the difference between the spectator approach presented so far and the alternative approach given by Becher and Leutwyler in Ref.~\cite{Bec99}. Instead of keeping the positive-energy pole of the \textit{heavy} particle, Becher and Leutwyler have rather kept the positive-energy pole of the \textit{light} particle. The corresponding contribution is most easily obtained by simply interchanging $A_1$ with $A_2$ on the right hand sides of Eqs.~(\ref{eq:A1A2split}) and~(\ref{eq:A1A2split2}), which corresponds to interchanging the masses $m$ and $M$ and the momenta $k$ and $P-k$. The relevant term is then given by 
\begin{eqnarray}\label{eq:SigmaBL}
\int_0^\infty  \frac{\mathrm dy}{(A_1y+A_2(1-y))^2}=\int_{-\infty}^1  \frac{\mathrm dz}{(A_2z+A_1(1-z))^2}\,,
\end{eqnarray}
where we have transformed the integral from a $y$ integration to a $z=1-y$ integration. This term leads to the desired contribution provided the limit $\epsilon_2\rightarrow0$ is taken \textit{before} the limit $\epsilon_1\rightarrow0$ and $W>0$. This \textit{alternative} \lq$\epsilon$-limiting' prescription contrasts the one from the CST contribution and ensures that there is only one pole in the lower $k_0$ plane, namely the positive-energy pole of the light particle. With this prescription, the imaginary part in the square brackets of Eq.~(\ref{eq:C}) changes the sign in the interval $-\infty
<z<1$ at $z=0$. Further, the interval $-\infty<z<0$ covered by the positive branch $z_+(s')$ is mapped onto the interval $0>s'(z)>-\infty$. This effectively results in a \lq$+\mathrm i \epsilon$' prescription in the interval $s_+<s<\infty$ and a \lq$-\mathrm i \epsilon$' prescription in the interval $-\infty<s<0$. Integrating Eq.~(\ref{eq:SigmaBL}) over $\mathrm d^2 k$ and then performing the $z$ integration yields the analytic result
\begin{eqnarray}\label{eq:BLeut1}
\Sigma_\text{BL}(s)=-\frac{1}{4\pi}\int_{-\infty}^1  \frac{\mathrm dz}{C(z,s)}=
\begin{cases}-\frac{\mathrm i}{2\pi\rho(s-\mathrm i \epsilon)}\left\lbrace\frac\pi2+\mathrm{arctan}\left[\eta_-(s-\mathrm i \epsilon)\right] \right\rbrace\,,
& -\infty<s<0\\
-\frac{\mathrm i}{2\pi\rho(s)}\left\lbrace\frac\pi2+\mathrm{arctan}\left[\eta_-(s)\right] \right\rbrace\,,
& 0<s<s_+\\
-\frac{\mathrm i}{2\pi\rho(s+\mathrm i \epsilon)}\left\lbrace\frac\pi2+\mathrm{arctan}\left[\eta_-(s+\mathrm i \epsilon)\right] \right\rbrace\,,
& s_+<s<\infty\quad.  
\end{cases}
\end{eqnarray}
The cut structure of $\mathrm {Im}\Sigma_\text{BL}(s)$ is displayed in Fig.~\ref{fig:ImSigmaBL3D}.

As expected, by interchanging the masses $M$ and $m$ the analytic results for $\Sigma_\text{CST}(s)$ and $\Sigma_\text{BL}(s)$ are mapped onto each other, as can be seen explicitly by comparing the expressions in~(\ref{eq:Bspec4}) and~(\ref{eq:BLeut1}). Finally, we obtain the following dispersion relation for the self-energy in the Becher-Leutwyler approach:
\begin{eqnarray}\label{eq:BLeut2}
\Sigma_\text{BL}(s)=\frac{1}{2\pi}\left[\frac12\int_{-\infty}^0\frac{\mathrm d s'}{\rho(s')}\frac{1}{s'-s+\mathrm i \epsilon }-\int^\infty_{s_+}\frac{\mathrm d s'}{\rho(s')}\frac{1}{s'-s-\mathrm i \epsilon }\right]
\end{eqnarray}with
\begin{eqnarray}\label{eq:ImSigmaBL}
\mathrm {Im}\Sigma_\text{BL
}(s)= -\frac{1}{2\rho(s)}\left[\frac12\theta(-s)+\theta(s-s_+)\right]\,
\end{eqnarray}
which agrees numerically with the analytic result of Eq.~(\ref{eq:BLeut1}).

\begin{figure}\begin{center}

\begin{center}
\resizebox{1.0\textwidth}{!}{
\begin{tabular}{cc}
\includegraphics[width=0.5\textwidth]{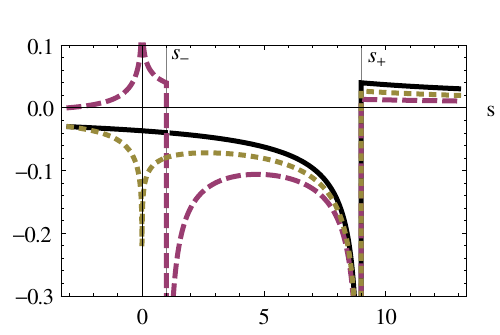}
&\includegraphics[width=0.5\textwidth]{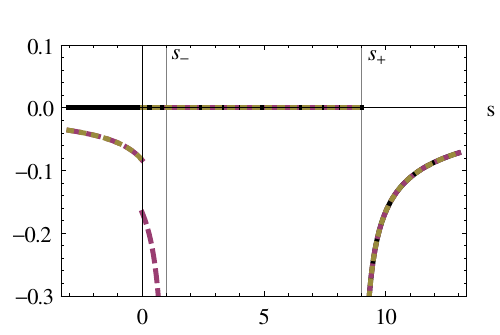}\\
\textbf{(a)}&\textbf{(b)}
\end{tabular}
}
\end{center}
\caption{The real parts (a) and imaginary parts (b) of $\Sigma_\text{CST}(s)$ (dashed line) and $\Sigma_\text{BL}(s)$ (dotted line) are compared with $\Sigma(s)$ (solid line) for the choice $M=2$ and $m=1$. Note that in (b) for $s<0$ $\mathrm {Im}\Sigma_\text{CST} (s)$ and $\mathrm {Im}\Sigma_\text{BL}(s)$ lie on top of each other and for $s>s_+$ all three lines lie on top of each other.}   \label{fig:Resigmacomp}
   \end{center}
\end{figure}

The real and imaginary parts of $\Sigma_\text{CST}(s)$ and $\Sigma_\text{BL}(s)$ are compared with the ones of $\Sigma(s)$ in Fig.~\ref{fig:Resigmacomp}.
Both $\Sigma_\text{CST}(s)$ and $\Sigma_\text{BL}(s)$ have left-hand cuts not present in the full amplitude. It is instructive to look at a related $\phi^2\chi^2$ theory to better understand the origin of these cuts.

\subsection{The \lq left-hand cut' in the two-particle scattering amplitude}
\label{sec:scatteringamplitude2D}

\begin{figure}[htbp]
	\begin{minipage}[b]{0.4\textwidth} 
	\includegraphics[width=1.1\textwidth]{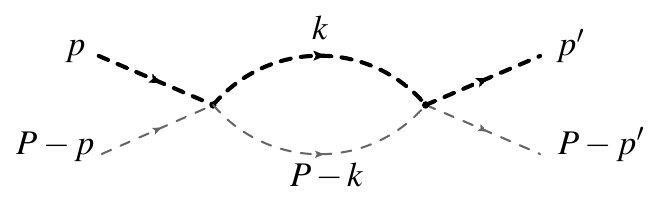}\vspace{1cm}
	\\\begin{center}
\textbf{(a)} 	  \end{center}
	\end{minipage}
	\hfill
	\begin{minipage}[b]{0.4\textwidth}
	\includegraphics[width=0.8\textwidth]{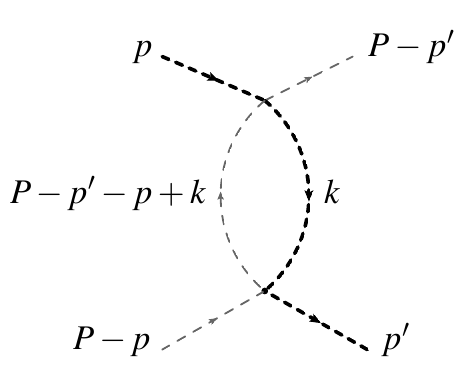}\\
	\begin{center}
\textbf{(b)}	\end{center}
	\end{minipage}
	\caption{The s-channel (a) and the u-channel (b) bubble diagram.}
   \label{fig:schannelbubble}
\end{figure}

From the topology of the self-energy graph in Fig.~\ref{fig:selfenergy} with one incoming and one outgoing line it is obvious that there exists only one channel, the s-channel we have discussed so far. If one considers, however, elastic scattering of two particles, then there are two topologically different Feynman diagrams that contribute to the second-order amplitude. These are the s- and the u-channel diagrams which are related by crossing symmetry. In this section we will show that for two-body scattering it is the u-channel contribution that gives rise to a \lq left-hand cut' along the real $s$-axis similar to the spectator \lq left-hand cut' discussed in the previous section. In order to see this explicitly we consider a $\phi^2\chi^2$-type theory in 1+1 dimensional Minkowski spacetime of two interacting scalar particles of masses $M$ and $m$. The corresponding second-order contributions to the scattering amplitude are the s- and the u-channel bubbles shown in Fig.~\ref{fig:schannelbubble}. 

The s-channel bubble contribution $B(s)$ of Fig.~\ref{fig:schannelbubble} (a) coincides with $\Sigma(s)$ of Eq.~(\ref{eq:B(s)1}) for the self-energy which has already been discussed in Sec.~\ref{sec:Disprel}, with the analytic result and the dispersion relation given by Eqs.~(\ref{eq:B(s)3}) and~(\ref{eq:B(s)5}), respectively. Therefore, we concentrate in this section on the u-channel bubble $B(u)$ of Fig.~\ref{fig:schannelbubble} (b). It is given by 

\begin{equation}\label{eq:Bu(u)1}
B(u)=\mathrm i\int \frac{\mathrm d^2 k}{(2\pi)^2}\frac{1}{A_1B_2}
\end{equation} with 
\begin{eqnarray}\label{eq:A1andB2}
B_2&=&m^2-(P-p'-p+k)^2-\mathrm i \epsilon_2
\end{eqnarray} 
and with Mandelstam $u=(P-p'-p)^2$. By introducing a Feynman parameter and performing the $\mathrm d^2k$ integration, this contribution can be put into a form that is equivalent to the one for $\Sigma(s)$, see Eq.~(\ref{eq:B(s)2}):
\begin{equation}\label{eq:Bu(u)2}
B(u)=-\frac{1}{4\pi}\int_0^1 \frac{\mathrm dz}{C(z,u)}\,.
\end{equation}
The fact that $B(u)$ and $B(s)$ have the same functional form is a consequence of crossing symmetry. Consequently,  $B(u)$ and $B(s)$ satisfy the same dispersion relations:
\begin{eqnarray}\label{eq:B(u)3}
B(u)=\frac{1}{\pi} \int_{s_+}^\infty \frac{\mathrm d s'\mathrm {Im} B(s')}{s'-u-\mathrm i \epsilon } 
\end{eqnarray}
with
\begin{eqnarray}\label{eq:ImBu}
\mathrm {Im}B(u)= -\frac{1}{2\rho(u)}\theta(u-s_+)\,.
\end{eqnarray}
If we consider only cases where Mandelstam $t$ is small, we can use the relation
\begin{eqnarray}\label{eq:relus}
s+u\approx 2M^2+2m^2
\end{eqnarray}
to express $u$ in terms of $s$, which gives us an approximate expression 
for the u-channel contribution as a function of Mandelstam $s$: 
\begin{eqnarray}\label{eq:B_u(s)}                                                                        B(u) \approx B(2M^2+2m^2-s)=B_u(s)\,.
                                                                         \end{eqnarray}
With this approximation the dispersion relation for $B_u(s)$ obtained from Eq.~(\ref{eq:B(u)3}) reads 
\begin{eqnarray}\label{eq:B_u(s)1}
B_u(s)=-\frac{1}{\pi} \int^{s_-}_{-\infty} \frac{\mathrm d s''\mathrm {Im} B_u(s'')}{s''-s+\mathrm i \epsilon }
\quad\text{with}\quad s''=2M^2+2m^2-s' 
\end{eqnarray}
and
\begin{eqnarray}\label{eq:ImBu2}
\mathrm {Im}B_u(s)= -\frac{1}{2\rho(s)}\theta(s_--s)\,.
\end{eqnarray}
We see that the u-channel bubble contribution expressed in terms of the Mandelstam $s$ variable has a \lq left hand cut' along the real $s$-axis going from $-\infty$ to $s_-$.  
With the help of Eq.~(\ref{eq:relus}) we can approximate the total second-order contribution as a function that depends solely on $s$:
\begin{eqnarray}\label{eq:totalM}
B(s)+B(u)\approx B(s)+B_u(s)=\mathcal M_2(s)\,.
\end{eqnarray} 
The CST and Becher-Leutwyler contributions to the s-channel bubble of Fig.~\ref{fig:schannelbubble}, $B_{\text{CST}}(s)$ and $B_{\text{BL}}(s)$, are simply the same as for the self-energy diagram of the preceding section, $\Sigma_{\text{CST}}(s)$ and $\Sigma_{\text{BL}}(s)$, respectively.

\begin{figure}\begin{center}
\begin{center}
\resizebox{1.0\textwidth}{!}{
\begin{tabular}{cc}
  $\alpha=1.05$&$\alpha=1.5$\\
  \includegraphics[width=0.5\textwidth]{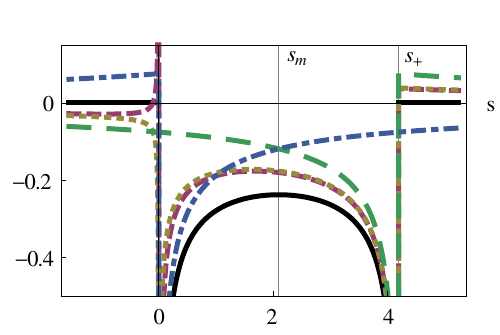} &  \includegraphics[width=0.5\textwidth]{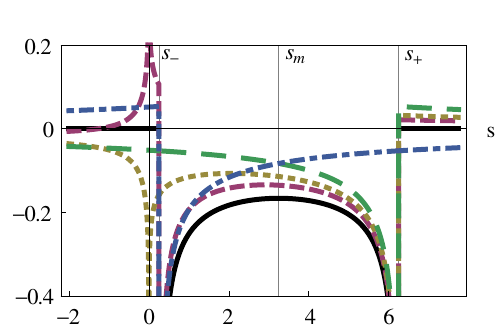}\\
\textbf{(a)} &\textbf{(b)}\vspace{1cm}\\
$\alpha=2$&$\alpha=9$\\
 \includegraphics[width=0.5\textwidth]{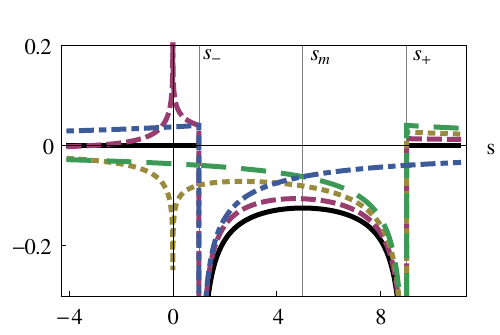}&
\includegraphics[width=0.51\textwidth]{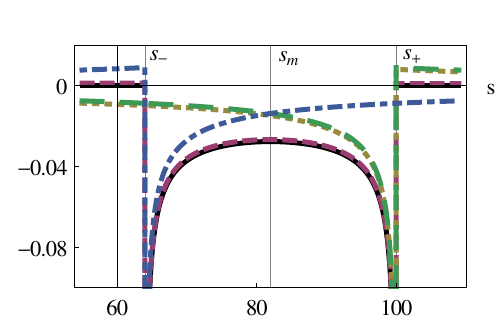}  \\
 \textbf{(c)}&\textbf{(d)}
\end{tabular}
}
\end{center}
\caption{The real parts $\mathrm{Re}\mathcal M_2(s)$ (solid), $\mathrm{Re}B_{\text{CST}}(s)$ (dashed), $\mathrm{Re}B_{\text{BL}}(s)$ (dotted), $\mathrm{Re}B(s)$ (long-dashed) and $\mathrm{Re}B_u(s)$} (dot-dashed) are shown for $m=1$ and for different choices $M=1.05$ (a), $M=1.5$ (b), $M=2$ (c) and $M=9$ (d).  
    \label{fig:CompareReBubble105}\end{center} 
\end{figure}

In Fig.~\ref{fig:CompareReBubble105} the real parts of $\mathcal M_2(s)$, $B_{\text{CST}}(s)$, $B_{\text{BL}}(s)$, $B(s)$ and $B_u(s)$ are shown for different mass ratios $\alpha=M/m$.
First, note the symmetry of the total contribution $\mathcal M_2(s)$ about the point $s=s_m\equiv M^2+m^2$ as consequence of crossing symmetry and $t\approx 0$. Second, from Fig.~\ref{fig:CompareReBubble105} (a), we observe that for the case of equal masses, i.e. $\alpha \simeq 1$, $B_{\text{CST}}(s)$ and $B_{\text{BL}}(s)$ become identical, as can be seen explicitly from the analytic expressions~(\ref{eq:Bspec4}) and ~(\ref{eq:BLeut1}), which differ only by interchanging the masses $M\leftrightarrow m$. Third, Fig.~\ref{fig:CompareReBubble105} (d) reveals that for the case when $M$ becomes very large compared to $m$, i.e. $\alpha \gg1$, the Becher-Leutwyler contribution $B_{\text{BL}}(s)$ goes over to the s-channel contribution $B(s)$ and the spectator contribution $B_{\text{CST}}(s)$ approaches the total contribution $\mathcal M_2(s)$. Actually, the latter observation is simply the confirmation of the CST cancellation theorem for $\phi^4$-type theories discussed in Ref.~\cite{Sta11}: in the limit $\alpha\rightarrow\infty$ 
the sum of the s- and u-channel diagrams becomes identical to the part of the s-channel diagram where the heavy particle is placed on its positive-energy mass shell. 
Therefore, we conclude that for mass ratios $\alpha \gg1$, the Becher-Leutwyler contribution is a better approximation to the self-energy diagram (or equivalently the s-channel bubble), whereas the CST contribution is a better effective description of the total bubble diagram (s- plus u-channel bubble). Consequently, for $\alpha \gg1$ the difference between the Becher-Leutwyler result and the CST result for the bubble diagram is the u-channel bubble contribution.                  

\section{Two scalar particles in 3+1 dimensions}
\label{sec:scalar3+1}
We now apply the above ideas to the  physical 3+1 dimensional Minkowski spacetime. In this case we are confronted with divergent integrals for the self-energy and for the bubble-diagram contributions. In order to find the dispersion relations for such diagrams we have to separate the regular from the singular parts. This will be achieved by a procedure~\cite{Bec99} based on dimensional regularization. 
\subsection{Dimensional regularization and dispersion relation}
\label{sec:dimregularization}
Let us first analyze the scalar-loop integral for the self-energy of Fig.~\ref{fig:selfenergy} in $d$ spacetime dimensions (with $d$ being an arbitrary value) and then perform the limit $d\rightarrow4$:
\begin{equation}\label{eq:Sigmad(s)1}
\tilde{\Sigma}(s)=\mathrm i\int \frac{\mathrm d^d k}{(2\pi)^d}\frac{1}{A_1A_2}\,.
\end{equation}
Here all two-vectors in $A_1$ and $A_2$ of Eq.~(\ref{eq:A1andA2}) have been replaced by  $d$-vectors.\footnote{As in Sec.\ref{sec:scalar1+1}, we use the standard convention for the metric tensor $\mathrm{diag} (\mathrm g_{\mu\nu})=(1,- 1,\ldots,-1)$ with $\mu,\nu=0,\ldots,d-1$. } Introducing one Feynman parameter and performing the $\mathrm d^dk$ integrations using the standard techniques leads to
\begin{equation}\label{eq:Sigmad(s)2}
\tilde{\Sigma}(s)=-\Gamma\left(2-\frac d2\right)\int_0^1 \frac{\mathrm dz}{(4\pi)^{\frac d2}}C^{\frac d2-2}(z,s)\,.
\end{equation}
This expression is singular for $d=4$ due to the $\Gamma$-function in front of the integral which reflects the logarithmic divergence of the loop integral~(\ref{eq:Sigmad(s)1}) in four dimensions. In order to separate the finite from the singular parts of $\tilde {\Sigma}(s)$ we write $d=4-\delta$ for $\delta>0$. Then, by expanding $\left[C(z,s)\right]^{-\frac\delta2}$ in~(\ref{eq:Sigmad(s)2}) about small values of $\delta$ we obtain
\begin{equation}\label{eq:Sigmadelta(s)3}
\tilde{\Sigma}_\delta(s)=-\frac{2\Gamma\left(1+\frac \delta2\right)}{\delta(4\pi)^{2-\frac \delta2}}\left[1-\frac\delta2\int_0^1 \mathrm dz\,  \mathrm {ln}\,C(z,s)\right]\,.
\end{equation}
Here the first term in the square brackets is identified as the singular part as $\delta\rightarrow0$. Since this part is independent of $s$ we define the finite part of $\tilde{\Sigma}(s)$ in $d=4$ dimensions by a single subtraction at a point $s=s_0$:
\begin{equation}\label{eq:barSigma(s)3}
\overline{\Sigma}(s)=\lim_{\delta\rightarrow 0}\left[\tilde{\Sigma}_\delta(s)-\tilde{\Sigma}_\delta(s_0)\right]=\frac{1}{(4\pi)^2}\int_0^1 \mathrm dz\,  \mathrm {ln}\left[\frac{C(z,s)}{C(z,s_0)}\right]\,.
\end{equation}
This integral is elementary and can be expressed as 
\begin{eqnarray}\label{eq:barSigma(s)5}
\overline{\Sigma}(s)&=&\frac{1}{(4\pi)^2}\left\lbrace\frac{\mathrm i \rho(s+\mathrm i \epsilon)}{s+\mathrm i \epsilon}\left\lbrace\mathrm {arctan}\left[\eta_+(s+\mathrm i \epsilon)\right]+\mathrm {arctan}\left[\eta_-(s+\mathrm i \epsilon)\right] \right\rbrace \right.\nonumber\\ &&-\left.\frac{\mathrm i \rho(s_0+\mathrm i \epsilon)}{s_0+\mathrm i \epsilon}\left\lbrace\mathrm {arctan}\left[\eta_+(s_0+\mathrm i \epsilon)\right]+\mathrm {arctan}\left[\eta_-(s_0+\mathrm i \epsilon)\right] \right\rbrace \right.\nonumber\\ &&+\left.
\mathrm{ln}\left(\frac Mm\right)^2\left[z_0(s+\mathrm i \epsilon)-z_0(s_0+\mathrm i \epsilon)\right]\right\rbrace\, ,
\end{eqnarray}
with
\begin{eqnarray}
 z_0(s)=\frac{1}{2s}\left(s+M^2-m^2\right)\,.
\end{eqnarray}
Note that the \lq$+\mathrm i \epsilon$' prescription again ensures that the imaginary part $\mathrm{Im} \overline{\Sigma}(s)$ is always evaluated on the correct side of the branch cut where it has a discontinuity, which is in this case always on the upper-half complex $s$ plane. 
The subtracted dispersion relation for $\tilde{\Sigma}(s)$ follows from Eq.~(\ref{eq:barSigma(s)3}) by an integration by parts and by transforming the integral from $z$ to $s'(z)$ (see the discussion in Sec.~\ref{sec:Disprel}):
\begin{eqnarray}\label{eq:tildeSigma(s)6}
\tilde{\Sigma}(s)=\tilde{\Sigma}(s_0)-\frac{s-s_0}{\pi}\int_{s_+}^\infty \frac{\mathrm d s'\,\mathrm {Im}\overline{\Sigma}(s')}{(s'-s_0)(s'-s-\mathrm i\epsilon)}\,
\end{eqnarray}
with
\begin{eqnarray}\label{eq:ImbarSigma6}
\mathrm {Im}\overline{\Sigma}(s)=-\frac{\rho(s)}{16 \pi s}\theta(s-s_+)\,.
\end{eqnarray}

This dispersion relation resembles the corresponding one in two dimensions, see Eq.~(\ref{eq:B(s)5}), with a branch-cut discontinuity along the real $s$ axis from $s_+$ to $\infty$.

\subsection{Spectator dispersion relation}
The spectator contribution $\tilde{\Sigma}_{\mathrm{CST}}(s)$ to the self-energy diagram in $d$ dimensions is obtained from $\tilde{\Sigma}(s)$ in the same manner as in Sec.~\ref{sec:speccontribution} by splitting up the product of the propagators according to Eq.~(\ref{eq:A1A2split2}). This gives

\begin{equation}\label{eq:Sigmaspec(s)1}
\tilde{\Sigma}_{\mathrm{CST}}(s)=-\Gamma\left(2-\frac d2\right)\int_0^\infty \frac{\mathrm dz}{(4\pi)^{\frac d2}}C^{\frac d2-2}(z,s)\,.
\end{equation}
Similar to $\tilde{\Sigma}(s)$ of the previous section this contribution is singular in $d=4$ dimensions because of the $\Gamma$-function in front of the integral. Furthermore, for $d=4$, the $z$ integral does not converge anymore due to the upper limit at $z=\infty$, which makes the CST case somewhat more intricate as compared to $\tilde{\Sigma}(s)$ from before. In order to work out the finite part of $\tilde{\Sigma}_{\mathrm{CST}}(s)$ we follow the procedure presented in Ref.~\cite{Bec99}. To this end we note that $C(z,s)$ is a second-order polynomial in $z$ that can be written in the form
\begin{equation}\label{eq:C(z,s)}
C(z,s)=C_0(s)+s\left[z-z_0(s)\right]^2
\end{equation}
where (suppressing the small imaginary parts)
\begin{eqnarray}
 C_0(s)=-\frac{\rho^2(s)}{4s}\,.
\end{eqnarray}Then
\begin{eqnarray}
 C^{\frac d2-2}(z,s)=C_0(s)C^{\frac d2-3}(z,s)+s\left[z-z_0(s)\right]^2C^{\frac d2-3}(z,s)
\end{eqnarray}
where the second term is proportional to the derivative of $C^{\frac d2-2}(z,s)$ with respect to $z$. An integration by parts gives 
\begin{equation}\label{eq:Sigmaspec(s)2}
\tilde{\Sigma}_{\mathrm{CST}}(s)=\frac{1}{(4\pi)^{\frac d2}}\frac{2\Gamma\left(3-\frac d2\right)}{(d-3)(d-4)}
\left\lbrace\left.\left[ z-z_0(s)\right]C^{\frac d2-2}(z,s)\right\rvert_{z=0}^{\infty}+(d-4)C_0(s)\int_0^\infty \mathrm dz\,C^{\frac d2-3}(z,s)\right\rbrace
\,.
\end{equation}
The first terms in the curly brackets are singular for $d=4$. Let us concentrate on the remaining finite term  of the integral. In four dimensions this integral converges and gives the finite part of the CST contribution (for $s\neq0$):
\begin{equation}\label{eq:barSigmaspec(s)1}
\overline{\Sigma}_{\mathrm{CST}}(s)=\frac{2\,C_0(s)}{(4\pi)^{2}}
\int_0^\infty \frac{\mathrm dz}{C(z,s)}= \begin{cases}-\frac{\mathrm i\rho(s-\mathrm i \epsilon)}{(4\pi)^2s}\left\lbrace\frac\pi2+\mathrm{arctan}\left[\eta_+(s-\mathrm i \epsilon)\right] \right\rbrace\,,
& -\infty<s<s_-\\
-\frac{\mathrm i\rho(s)}{(4\pi)^2s}\left\lbrace\frac\pi2+\mathrm{arctan}\left[\eta_+(s)\right] \right\rbrace\,,
& s_-<s<s_+\\
-\frac{\mathrm i\rho(s+\mathrm i \epsilon)}{(4\pi)^2s}\left\lbrace\frac\pi2+\mathrm{arctan}\left[\eta_+(s+\mathrm i \epsilon)\right] \right\rbrace\,,
& s_+<s<\infty\quad.  
\end{cases}
\end{equation}
Note that this function has a pole at $s=0$. Following ~\cite{Bec99}, this suggests the twice-subtracted dispersion relation
\begin{eqnarray}\label{eq:tildeSigmaspec}
\tilde\Sigma_\text{CST}(s)=\frac{z_0(s)}{z_0(s_+)}\tilde\Sigma_\text{CST}(s_+)+
\frac{\rho^2(s)}{(4\pi)^2s}\left\lbrace\left[\frac12\int_{-\infty}^0+\int_{0}^{s_-}\right]\frac{\mathrm d s'}{\rho(s')}\frac{1}{s'-s+\mathrm i \epsilon }-\int^\infty_{s_+}\frac{\mathrm d s'}{\rho(s')}\frac{1}{s'-s-\mathrm i \epsilon }\right\rbrace \, ,
\end{eqnarray}from which the imaginary part follows as
\begin{eqnarray}\label{eq:ImtildeSigmaSpec}
\mathrm {Im}\overline{\Sigma}_\text{CST}(s)= -\frac{\rho(s)}{16\pi s}\left[\frac12\theta(-s)+ \theta(s)\theta(s_--s)+\theta(s-s_+)\right]\,.
\end{eqnarray}
Thus, we encounter the typical CST branch-cut structure of the two-dimensional case, see Eq.~(\ref{eq:Bspec5}).

The Becher-Leutwyler contribution $\tilde\Sigma_\text{BL}(s)$ to the self-energy diagram in four dimensions is expressed by similar formulae (see Sec.~\ref{sec:speccontribution}) and will not be given here explicitly.   
\subsection{Bubble diagram}
\begin{figure}\begin{center}

  \includegraphics[width=0.6\textwidth]{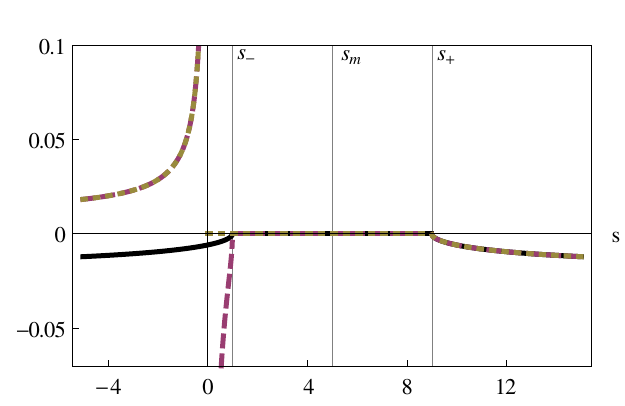}
\caption{The imaginary parts $\mathrm{Im}\overline{\mathcal M}_2(s)$ (solid), $\mathrm{Im}\overline{B}_{\text{CST}}(s)$ (dashed) and $\mathrm{Im}\overline B_{\text{BL}}(s)$ (dotted) are shown for the choice $M=2$ and $m=1$. Note that for $s>s_-$ all three curves, and for $s<0$  $\mathrm{Im}\overline{B}_{\text{CST}}(s)$ and $\mathrm{Im}\overline{B}_{\text{BL}}(s)$, lie on top of each other.}
 \label{fig:CompareImBubblefourDim} 
   \end{center}
\end{figure}

The contributions to the s-channel bubble diagram (Fig.~\ref{fig:schannelbubble} (a)) in four dimensions, $\tilde B(s)$, $\tilde B_\text{CST}(s)$ and $\tilde B_\text{BL}(s)$ coincide with the ones for the self-energy, $\tilde \Sigma(s)$, $\tilde \Sigma_\text{CST}(s)$ and $\tilde \Sigma_\text{BL}(s)$, respectively. What is still missing for the calculation of the full bubble diagram in four dimensions is the contribution of the $u$-channel bubble diagram of Fig.~\ref{fig:schannelbubble} (b). In the approximation of Eq.~(\ref{eq:B_u(s)}) this contribution can be written as a function of Mandelstam $s$ only. Using the methods of Sec.~\ref{sec:dimregularization} this contribution satisfies the following once-subtracted dispersion relation: 
\begin{eqnarray}\label{eq:tildeBu(s)}
\tilde{B}_u(s)&=&\tilde{B}_u(s_0)+\overline{B}_u(s)\nonumber\\&=&\tilde{B}_u(s_0)-\frac{s-s_0}{\pi}\int_{-\infty}^{s_-} \frac{\mathrm d s'\,\mathrm {Im}\overline{B}_u(s')}{(s'-s_0)(s'-s+\mathrm i\epsilon)}\,
\end{eqnarray}
with
\begin{eqnarray}\label{eq:ImbarBu}
\mathrm {Im}\overline{B}_u(s)=-\frac{\rho(s)}{16 \pi (2M^2+2m^2-s)}\theta(s_--s)\,.
\end{eqnarray}
As expected from the result in two dimensions, we see that also in four dimensions the $u$-channel bubble contribution gives raise to a \lq left-hand branch cut' in the complex $s$-plane along the real axis from $s=-\infty$ to $s=s_-$. 
Finally, we can write down the approximate expression for the finite part of the total second-order bubble amplitude:
\begin{eqnarray}\label{eq:tildeM2(s)}
\overline{\mathcal M}_2(s)&=\overline{B}(s)+\overline{B}_u(s)\,.
\end{eqnarray}
The imaginary parts of the contributions are compared in Fig.~\ref{fig:CompareImBubblefourDim}.

\begin{figure}

\begin{center}
\resizebox{1.0\textwidth}{!}{
\begin{tabular}{cc}
$\alpha=1.1$&$\alpha=2$\\
 \includegraphics[width=0.5\textwidth]{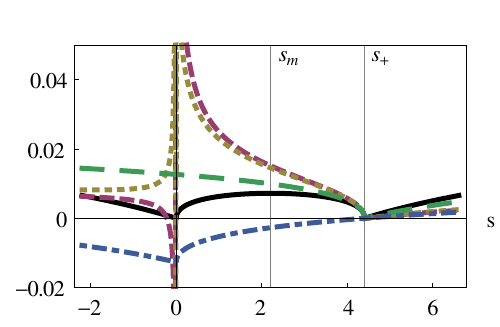} & \includegraphics[width=0.5\textwidth]{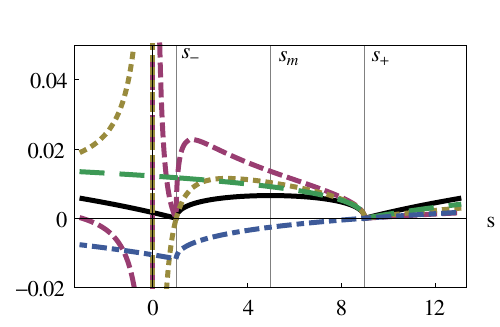}\\
 \textbf{(a)}&\textbf{(b)}\vspace{1cm}\\
 $\alpha=5$&$\alpha=50$\\
 \includegraphics[width=0.5\textwidth]{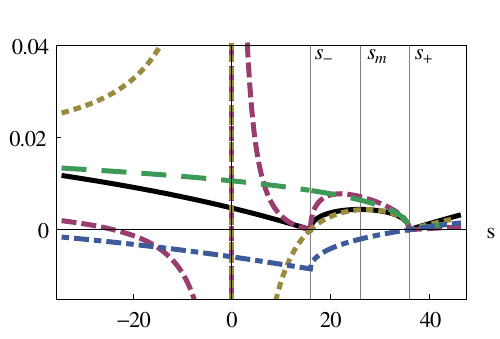}& \includegraphics[width=0.53\textwidth]{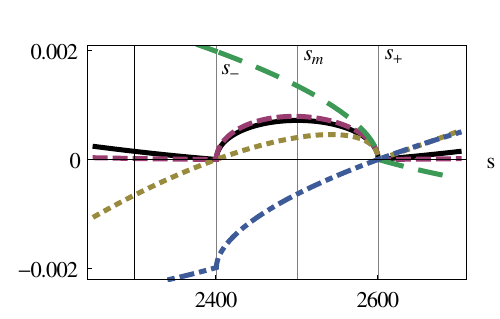}\\
 \textbf{(c)}&\textbf{(d)} 
 \end{tabular}
 }
\end{center}

 
\caption{The real parts $\mathrm{Re}\overline{\mathcal M}_2(s)$ (solid), $\mathrm{Re}\overline{B}_{\text{CST}}(s)$ (dashed), $\mathrm{Re}\overline B_{\text{BL}}(s)$ (dotted), $\mathrm{Re}\overline B(s)$ (long-dashed) and $\mathrm{Re}\overline B_u(s)$} (dot-dashed) are shown are shown for $m=1$ and for different choices $M=1.1$ (a), $M=2$ (b), $M=5$ (c) and $M=50$ (d).   
\label{fig:CompareReBubblefourDim11} 
\end{figure}

It is interesting to see the difference between the imaginary parts in two and in four dimensions. While the former are singular at $s=s_{\pm}$ (due to the $\rho$ function in the denominator), the latter, in contrast, vanish at $s=s_{\pm}$ ($\rho$ function in numerator). Furthermore, in four dimensions the imaginary parts of $\overline B_{\text{CST}}(s)$  and $\overline B_{\text{BL}}(s)$ are singular at $s=0$ and become positive for $s<0$, whereas in two dimensions they are always negative.  

In order to compare the real parts we choose the subtraction point for $\overline{B}(s)$ and $\overline{B}_u(s)$ to be at threshold, i.e. $s_0=s_+$.
The real parts of the contributions are shown in Fig.~\ref{fig:CompareReBubblefourDim11} for different mass ratios $\alpha=\frac Mm$. 
As in two dimensions, the symmetry of $\overline{\mathcal M}_2(s)$ about the point $s_m=M^2+m^2$ is a consequence of crossing symmetry. Further, as expected from the symmetric appearance of the masses in the expressions, for equal masses ($\alpha\simeq1$)  $\mathrm{Re}\overline{B}_{\text{CST}}(s)$ and $\mathrm{Re}\overline B_{\text{BL}}(s)$ become identical. Also, for $\alpha\rightarrow\infty$ the CST result approaches the total bubble contribution.  This happens, however, not as fast as in two dimensions. Nevertheless, we can still conclude that the CST result approximates the sum of the two bubble diagrams quite well for $M\gg m$. Interestingly, in four dimensions the Becher-Leutwyler result does not converge to the total self-energy contribution as $\alpha\rightarrow\infty$ and is therefore not  as good an approximation in this limit as it is in two dimensions. 

\section{Confining interaction between scalar particles}
\label{sec:scalar-conf}
We consider now a simplified model of mesons as bound states of scalar particles in the framework of CST, whose interaction consists of a confining part and another one that is described through the exchange of a second scalar particle. In order to make this description consistent, the self-energy of these scalar particles, which we will call simply ``scalar quarks'', has to be calculated from the same interaction kernel. The dressing by the scalar exchange has already been calculated in Secs.\ \ref{sec:scalar1+1} and \ref{sec:scalar3+1}, so we only need to calculate the contribution from the confining part.

Following the general form given in Eqs.\ (\ref{eq:qqbar-kernel}), (\ref{eq:VL}), and (\ref{eq:VA}), a relativistic generalization in CST can be achieved by \cite{Gross:1991te}
\begin{equation}
V_{\mathrm{RL}}(\hat{p},\hat{k}) = \lim_{\epsilon \rightarrow 0} \left[ V_{\mathrm{RA}}(\hat{p},\hat{k}) -2E_k (2\pi)^3 \delta^{(3)}({\bm p}-{\bm k})\int \frac{\mathrm d^3 k'}{2E_{k'} (2\pi)^3} V_{\mathrm{RA}}(\hat{p},\hat{k}') \right] \, ,
\label{eq:VRL}
\end{equation}
and 
\begin{equation}
V_{\mathrm{RA}}(\hat{p},\hat{k}) = - \frac{8\pi\sigma}{(-q^2+\epsilon^2)^2} \, ,
\label{eq:VRA}
\end{equation}
with 
\begin{equation}
q^2=\left( E_p-E_k\right)^2-({\bm p}-{\bm k})^2 \, .
\end{equation}
The expression for the self-energy of a scalar quark dressed by $V_{\mathrm{RL}}$ in 3+1 dimensions is then simply
\begin{equation}
\label{eq:Sigma-L}
\Sigma_\mathrm{L,CST}(s) = \int \frac{\mathrm d^3 k}{2E_{k} (2\pi)^3}V_{\mathrm{RL}}(\hat{p},\hat{k}) = 0 \, ,
\end{equation}
which is easy to verify from the form of the kernel (\ref{eq:VRL}). That the integral of $V_{\mathrm{RL}}$ should vanish follows from the fact that $V_\mathrm{L}({\bm q})$ of Eq.\ (\ref{eq:VL}) is the Fourier transform of the linear potential $\tilde{V}(r)=\sigma r$, and the inverse Fourier transform must yield zero at $r=0$. This yields the condition
\begin{equation}
\tilde{V}(0)=\int \frac{\mathrm d^3 q}{(2\pi)^3} V_\mathrm{L}({\bm q})=0 \, ,
\label{eq:V(0)}
\end{equation}
and Eq.\ (\ref{eq:Sigma-L}) can then be seen as a covariant generalization of (\ref{eq:V(0)}).

To summarize, we have shown that the linear confining interaction does not contribute to the scalar-quark self-energy at all. If the total kernel consists of a confining part and another scalar exchange as described in Sec.\ \ref{sec:scalar3+1}, then the total result for the self-energy is given by Eq.\ (\ref{eq:tildeSigmaspec}).
While the scalar confinement makes no contribution to the self-energy, it does contribute to the structure of the meson.  This can be understood physically if we think of the self-energy as a correction that occurs at short distances (i.e. at ${\bm r}=0$), while the meson wave function is sensitive to the kernel at all values of ${\bm r}$. 

For completeness it should be mentioned that in 1+1 dimensions the same result is obtained.

\section{Summary and Conclusions}
\label{sec:conclusions}
In this paper, we continue the development of a unified and consistent covariant quark model for mesons in the framework of the CST that was initiated by Gross and Milana \cite{Gross:1991te,Gross:1991pk,Gross:1994he}. The main aspect of this model we aim to improve is the consistent implementation of the quark mass function, in the sense that it is calculated from the same interaction that binds the two quarks together. In a first step, we investigated the simpler case of scalar quarks and calculated their self-energy amplitude, obtaining a representation for the mass function in form of a dispersion integral. The CST approach modifies the analytic structure of field-theoretic amplitudes by selecting certain pole contributions and neglecting others. It is therefore interesting to see that the CST amplitudes actually do satisfy dispersion relations. 

To avoid complications due to divergences in loop integrals, we performed these calculations first in 1+1 dimensions. 
We found that the CST amplitude satisfies a dispersion relation which has a left-hand cut, in addition to the usual right-hand cut. This is similar to an approximation investigated by Becher and Leutwyler, which also has a left-hand cut. In order to understand the origin of the left-hand cut in the CST amplitude, we calculated also the dispersion relations for the scattering amplitude of two scalar particles in a $\phi^2\chi^2$-type theory to second order. The calculation in the s-channel is essentially the same since only the external particle momentum has to be written as the sum of two particle momenta, and therefore the  same cut structure is obtained. However, there is also a u-channel which generates a left-hand cut. Thus, the total amplitude has both a left-hand and a right-hand cut, just as the s-channel CST amplitude with the heavier particle on its positive-energy mass shell. This is a natural result, since one of the basic ideas of the CST 
is that its amplitudes approximate the sum of direct and crossed diagrams, which in this case means s- and u-channels. On the other hand, in the case of the self-energy amplitude at one-loop level, there is no u-channel, and therefore a simple physical interpretation for the presence of a left-hand cut is not obvious. At this time the interpretation and treatment of this left-hand cut is an open question to be discussed in future work.
Repeating these calculations in 3+1 dimensions using dimensional regularization gave the same cut structure for the self-energy. 

Because of the CST constraint (\ref{eq:Sigma-L}) on the confining interaction, the self-energy of a scalar quark due to a linear confining interaction vanishes exactly, and does not contribute to the mass function at all. The confining interaction will contribute to the structure of the meson because the meson wave function is sensitive to the kernel at all values of ${\bm r}$ (or $q$) while the self-energy is a correction that occurs only at short distances where a purely linear confining interaction is zero. It would be amusing to study a simple toy model of scalar particles bound solely through a  linear confining interaction where a constant mass for the constituent particles would be part of an exact and consistent solution of the system.  We postpone discussion of this toy model for another time.

 We  emphasize that the vanishing of the self-energy contribution due to confinement holds only for scalar particles and does not carry over to spinor quarks. With spinor quarks the structure of the confining interaction can be divided into a part that gives no contribution to the self energy but may violate chiral symmetry (and therefore decouples in the chiral limit) and a chirally invariant part that will, in general, contribute to the quark self-energy.  These issues will be discussed when we extend these studies to the case of spin-1/2 quarks with more realistic interactions, including  confining interactions with a spin dependent structure, and when we consistently calculate the mass function, the meson spectrum, and the meson wave functions.
 
\begin{acknowledgements}
This work received financial support from Funda\c c\~ao para a Ci\^encia e a Tecnologia (FCT) under grant Nos.~PTDC/FIS/113940/2009 and POCTI/ISFL/2/275. This work was also partially supported by the European Union under the HadronPhysics3 Grant No. 283286, and by Jefferson Science Associates, LLC under U.S. DOE Contract No. DE-AC05-06OR23177.
\end{acknowledgements}

\vspace{-3mm}
\bibliographystyle{spmpsci}      
\bibliography{disrel-v4.3}

%
%

\end{document}